\documentclass[12pt]{iopart}
\usepackage{amsfonts}
\usepackage{amssymb}
\usepackage{epsfig}
\usepackage{graphicx}
\usepackage{subfigure}

\usepackage{color}

\def\black{\color{black}}

\def\red#1{\color{red}#1\black}

\def\beq{\begin{equation}}
\def\eeq{\end{equation}}
\def\beqa{\begin{eqnarray}}
\def\eeqa{\end{eqnarray}}
\def\hf{\textstyle{1\over2}}
\def\half{{1\over2}}

\begin{document}

\title[SHA and asymptotic SU(2) and SU(1,1) CG coefficients]{The shifted harmonic approximation and asymptotic SU(2) and SU(1,1) Clebsch--Gordan coefficients\\
\rm (published in J. Phys. A: Math. Theor. 43 (2010) 505307)}

\author{D.J.~Rowe$^\dag$ and Hubert de Guise$^\ddag$ }

\address {$^\dag$Department of Physics, University of Toronto, Toronto, ON M5S 1A7, Canada\\
\hspace{1cm}  E-mail:  rowe@physics.utoronto.ca}

\address{$^\ddag$Department of Physics, Lakehead University,
Thunder Bay, ON, P7B 5E1, Canada}

\date{\today}

\begin{abstract}
Clebsch-Gordan coefficients of SU(2) and SU(1,1) are defined as eigenfunctions of a linear operator acting on the tensor product of the Hilbert spaces for two irreps of these groups.  The  \emph{shifted harmonic approximation}  is then used to solve these equations in asymptotic limits in which these eigenfunctions approach harmonic oscillator wave functions and thereby derive asymptotic expressions for these  Clebsch--Gordan coefficients.
\end{abstract}



\section{Introduction}

The shifted harmonic approximation (SHA) is an approximation to a technique for realizing a set of operators, defined initially as linear transformations of a finite vector space, as differential operators.  It was introduced for the purpose of understanding the nature of phase transitions in systems with su(2) or
su(2) $\oplus$ su(2) spectrum generating algebras \cite{Chen} and subsequently applied to a model with an su(1,1) $\oplus$ su(1,1) spectrum generating algebra \cite{SHAIBM}. More recently it has been developed for application to a multi-level pairing model \cite{SYHo}, for which the spectrum generating algebra is a direct sum of multiple su(2) algebras.  It can be regarded
as a technique for
contracting a Lie algebra representation to that of a simpler algebra in well-defined limiting situations.

In this paper, we show how the SHA provides approximate expressions for SU(2) and SU(1,1) Clebsch-Gordan coefficients that  become precise in certain asymptotic limits in which they approach harmonic oscillator wave functions when regarded as functions of appropriate parameters.
Moreover, we show by examples, that these limits are approached very rapidly and provide remarkably accurate approximations more generally.

In addition to the examples given above,
the groups SU(2) and SU(1,1) and the coupling of their irreducible
representations have numerous applications in quantum optics \cite{qopticalbooks}.%
\footnote{A highly non-exhaustive list contains, for instance  \cite{qopticalapplications}.}  In this context, 
asymptotic SU(2) and SU(1,1) CG coefficients are of huge potential value in
problems  in which a prohibitive amount of time is often needed to compute the required coefficients numerically from the known analytical formulae.
In contrast, the asymptotic coefficients given here can be calculated in fractions of a second.  Asymptotic SU(2) CG coefficients have also been connected to a tight-binding model of a one-dimensional potential \cite{Sprung}.

Clesbch--Gordan (CG) coefficients $( s_{1},M-m,\, s_{2},m|SM )$ for the group SU(2)
are required when two systems of spin
$\vec{s}_{1}$ and $\vec{s}_{2} $ are coupled to total spin $\vec{S}$:
\begin{equation}
|SM\rangle =\sum_{m} |s_{2}m\rangle\otimes |s_{1},M-m\rangle
(s_{1},M-m,\,s_{2},m|SM).  \label{Sstates}
\end{equation}
In Eqn.\ (\ref{Sstates}), the projections along the common quantization axis of the
spins $\vec{s}_{1},\vec{s}_{2}$, and $\vec{S}$ are $M-m,m$ and $M$, respectively.
Eqn.\ (\ref{Sstates}) is also applicable to a wide variety of other quasi-spin systems having
states that carry su(2) irreps.

Because of the connection between SU(2) and SO(3), CG
coefficients also appear in problems where products of spherical
harmonics and other related special functions occur naturally.
Furthermore, the ubiquity of su(2) as a subalgebra of other Lie
algebras makes the SU(2) CG coefficients (and those of  SU(1,1) for non-compact groups)
an important ingredient
in the computation of coupling coefficients for higher groups.

Much is known about SU(2) CG coefficients: any coefficient can be
obtained in closed form using an expression containing
a sum of square roots of rational factors. However,
for small values of $m$ and $M$, the complexity of this sum increases rapidly with 
$S$, and numerous asymptotic estimates for large $S$ have been developed to understand and quickly evaluate CG coefficients in this regime.

A first consideration of asymptotic limits for SU(2) coefficients
 is due to Wigner \cite{Wigner}.
He approached the problem from a semi--classical perspective and
obtained average expressions for the coefficients when all momenta
were large; unfortunately, Wigner's result did not capture the
essentially oscillatory nature of asymptotic CG coefficients and
left the door open for further studies.  Brussaard and Toelhoek
\cite{Brussaard} used the  WKB method to connect
reduced SU(2)--Wigner functions with asymptotic CG
coefficients.  Their result,
which applies to an asymptotic limit in which just
two angular momenta $s_1$ and $S$ are large, significantly improved on Wigner's
because it correctly gave the sign of the coefficient and turns out
to be reasonably accurate, even for modest values of $s_1$
and $S$.
Finally, Ponzano and Regge \cite{PonzanoRegge}, in their
authoritative work grounded almost entirely on geometrical arguments,
expanded on \cite{Wigner} and \cite{Brussaard} by providing accurate expressions for CG
coefficients valid for three large angular momenta.

The insights provided by \cite{Brussaard} and \cite{PonzanoRegge}
have been a source of inspiration for many subsequent authors. A detailed
survey of the literature pre-1988 can be found in
\cite{varshalovich}. Comparatively recent work, limited to asymptotic SU(2) CG
coefficients with all three momenta large, include, for instance,
Refs.\ \cite{Reinsch,Schulten}, wherein a detailed and systematic review of the results of Ponzano and Regge can be found, with emphasis on
closing loose ends in some
of their ``heuristic'' arguments.

In this paper we show that for asymptotically large values of $s_1$ and $s_2$,
and for finite values of $n$ and $M$, the SU(2) CG coefficients
rapidly approach the asymptotic expressions
\beqa
&&( s_{1},\!M\!-\!m,\; s_{2}m |s_{1}\!+\!s_{2}\!-n,M)\nonumber \\
&&\qquad \sim
(-1)^n \Big(\frac{a}{\sqrt{\pi} 2^n n!}\Big)^{\frac12}
H_n\big( a (m-x_0)\big) e^{-\frac12 a^2(m-x_0)^2} , \label{mainsu2result}
\eeqa
where $H_{n}$ is a Hermite polynomial,
with parameters defined by a simple SHA algorithm.
It is also shown that these asymptotic results have analytic  approximations which remain precise in the asymptotic limit but are approached somewhat less   rapidly; they are given by the explicitly expression
\beqa
&&( s_1 m_1\, s_2 m_2 |s_{1}\!+\!s_{2}\!-n,M)\nonumber \\
&&\qquad \sim
(-1)^n \Big(\frac{a}{\sqrt{\pi} 2^n n!}\Big)^{\frac12}
H_n\Big( a\, \frac{\sigma_1m_2 - \sigma_2m_1}{\sigma_1+\sigma_2}\Big) e^{-\frac12 a^2(m-x_0)^2} , \label{mainsu2result2}
\eeqa
with
\beqa
a^4 &=& \frac{(\sigma_1+\sigma_2)^4}
{\sigma_1^2\sigma_2^2 \big[ (\sigma_1+\sigma_2)^2-M^2\big]} \,, \qquad\qquad
x_0=\frac{\sigma_2 M}{\sigma_1+\sigma_2}\, ,\label{x0intro}\\
\sigma_1&=&\sqrt{s_{1}(s_{1}+1)}\,,\qquad \sigma_{2}=\sqrt{s_{2}(s_{2}+1)}\, .
\eeqa

Basis states for unitary  irreps of the positive discrete series of  su(1,1) within the tensor product of two such irreps are also given by linear combinations
\beq
\vert Kn\rangle = \sum_{n_1n_2} |k_2n_2\rangle \otimes |k_1n_1\rangle
\langle k_1k_2\,Nm |Kn \rangle,
\eeq
where $\langle k_1k_2\, Nm |Kn\rangle $ is an SU(1,1) CG coefficient related to a non-vanishing coefficient in the notation of Van der Jeugt \cite{VdJ}, by
\beq
\langle k_1k_2\, Nm |Kn\rangle = C(k_1,n_1, k_2,n_2; K,n) \label{su11cgintro}
\eeq
with  $N= n_1+n_2$ and $m= n_2-n_1$. Note that an SU(1,1) CG coefficient vanishes unless
 $K+n = k_1+n_1+k_2 +n_2= k_1+k_2+N$.
The notation is clarified in Sect.\ref{sec:su11theory}.

Various  closed form expressions for the SU(1,1) coefficients can be found in the literature.
The exact expressions of immediate relevance to our work are given, for instance, in \cite{VdJ,su11CGpapers}.
In addition, several other authors \cite{otherworksu11} have published expressions spanning a variety of couplings of representations (not only of two positive discrete series) and a number of different bases.

In spite of these exact results,
there is limited knowledge of the asymptotic behavior of the coefficients, 
although some basic results valid
for the coupling of two irreps of the positive discrete series with $k_1\to\infty$ and $k_2$ finite  can be found in \cite{CGHdGDJR}, a result that proves to be useful in constructing states in odd nuclei \cite{rotorsp3} within the context of the nuclear symplectic model.

The SHA approach is applied in Sect.\ \ref{sec:su11theory} to derive asymptotic expressions for the SU(1,1) CG coefficients of Eqn.(\ref{su11cgintro}) in an altogether different regime to that studied in \cite{rotorsp3}.
The result, similar to that given above for a class of SU(2) coefficients,  is that
\beqa \langle k_1k_2\, Nm |Kn \rangle  \nonumber\\
 \qquad\qquad\sim \displaystyle
(-1)^{N+m} \Big(\frac{a}{\sqrt{\pi} 2^n n!}\Big)^{\frac12}
H_n\big( a (m-x_0)\big) e^{-\frac12 a^2(m-x_0)^2} . \label{eq:ASSU11coefs}
\eeqa
when $N=n_1+n_2$ becomes asymptotical large and $n$ remains finite, and with parameters defined by a simple SHA algorithm given in Sect.\ \ref{sect:su11SHA}.
It is also shown that precise asymptotic expressions which, however, are approached somewhat less   rapidly, are given by Eqn.\ (\ref{eq:ASSU11coefs}) with the explicit parameter values
\beq
a^2 =  \frac{N+ 4\sqrt{\kappa_1\kappa_2}}
{2N\sqrt{\kappa_1\kappa_2}},\quad
x_0 = \frac{N(\kappa_1-\kappa_2)}{N+ 4\sqrt{\kappa_1\kappa_2}} . \label{eq:su11analytic}
\eeq
where
\beq
\kappa_1=k_1+\frac{N}{4}\,,\qquad \kappa_2=k_2+\frac{N}{4}\, .
\eeq

The SHA method described in this paper complements other approaches where the
asymptotic behavior is examined using WKB methods.
As mentioned briefly in the Discussion, the SHA technique can be regarded as a procedure for obtaining the contraction of a Lie algebra to a simpler Lie algebra that is appropriate in certain limiting situations.  Such contractions are known to lead to valuable insights and useful approximation procedures in physics as evidenced, for example, in the approach of quantum mechanics to classical mechanics as $\hbar$ or some other
scale parameter approaches zero, in the approximation of fermion pair algebras by boson algebras in the random-phase approximation of many-body theory, and in the bosonic behaviour of
large atomic samples \cite{cooling}.

\section{Asymptotic SU(2) Clebsch-Gordan coefficients}\label{sec:su2CG}

Let $\{\hat{S}_{+},\hat{S}_{-},\hat{S}_{0}\}$ satisfy the usual commutation relations of the complex extension of the su(2) Lie algebra:
\begin{equation}
\lbrack \hat{S}_{0},\hat{S}_{\pm }]=\pm \hat{S}_{\pm },\quad
\lbrack \hat{S}_{+},\hat{S}_{-}]=2\hat S_{0}.  \label{su21}
\end{equation}
Basis states, $|sm\rangle$, for an su(2) irrep are then defined by the equations
\beqa \hat S_0 |sm\rangle &=& m|sm\rangle, \\
    \hat S_\pm  |sm\rangle &=& \sqrt{(s\mp m)(s\pm m+1)}\, |s,m\pm 1\rangle.
\eeqa

\emph{Coupled} basis states for irreps of su(2) within the tensor product of two such irreps are  given by
\beq
|SM\rangle =
\sum_{m_1m_2} |s_2m_2\rangle \otimes |s_1m_1\rangle
( s_1m_1\, s_2m_2 |SM),
\label{coupledsu2}
\eeq
where $( s_1m_1\, s_2m_2 |SM )$ is an SU(2) CG coefficient.
These coefficients are equal to the overlaps of coupled and uncoupled tensor product states
\beq (s_1m_1\,s_2m_2|SM) =
 \big[ \langle s_1m_1|\otimes \langle s_2m_2| \big] |SM\rangle .
 \eeq

For convenience, we denote the tensor product states by
\beq|s_1s_2 Mm\rangle \equiv |s_2m_2\rangle \otimes |s_1m_1\rangle ,\label{eq:Mm}\eeq
with $M= m_1+m_2$ and $m=m_2$.
The SU(2) CG coefficients are then the overlaps
\beq
( s_1m_1\, s_2m_2 |SM )  =
  \langle s_1 s_2 Mm |SM\rangle =
 \big[ \langle s_1m_1|\otimes \langle s_2m_2| \big] |SM\rangle .
 \eeq
They are defined (to within arbitrary phase factors) by the requirement that the states $\{ |SM\rangle\}$ satisfy the eigenvalue
equations
\beqa
\hat S_0 |SM\rangle = M|SM\rangle, \\
 \hat S_+\hat S_-  |SM\rangle = \big[S (S+1)-  M(M-1)\big]\, |SM\rangle,
    \label{eq:su2action}
\eeqa
with
\beq
\hat S_0 = \hat S_0^1 + \hat S_0^2 , \quad
\hat S_\pm = \hat S_\pm^1 + \hat S_\pm^2 . \label{eq:sumS}
\eeq

\subsection{The shifted harmonic approximation} \label{sect:su2SHA}

We now determine the CG coefficients using the SHA and show them to be precise in the
limit of asymptotically large values of $s_1$, $s_2$, and $S$ and finite values of $M$
and $n= s_1+s_2-S$.

The desired CG coefficients are first expressed
as overlap functions, in the form
\beq
\psi _{n}^{s_{1}s_{2}M}(m)=\langle s_1s_2Mm |s_{1}\!+\!s_{2}\!-\!n,M\rangle.
\eeq
This notation is introduced with the intention that, for given values of $s_1$, $s_2$, $M$ and $n$,
a set of CG coefficients can be regarded as the values of a function, $\psi _{n}^{s_{1}s_{2}M}$,  of the discrete variable $m$.
Moreover, the state $|SM\rangle$ with
$S=s_{1}\!+\!s_{2}\!-\!n$ is completely determined by the values,
$\{ \psi _{n}^{s_{1}s_{2}M}(m), m= -s_2, \dots, s_2\}$, which implies that $\psi _{n}^{s_{1}s_{2}M}$ can be interpreted as a wave function for
the state  $|s_{1}\!+\!s_{2}\!-\!n,M\rangle$.

The operators $\hat S_0$, $\hat S_\pm$ are
now mapped to  operators on these wave
functions defined by
\beqa
\hat{\mathcal{S}}_\nu \psi _{n}^{s_{1}s_{2}M}(m)
&\equiv & \langle s_1s_2 Mm | \hat S_\nu |s_{1}\!+\!s_{2}\!-\!n,M \rangle \nonumber\\
&=&\sum_{M'p}   \langle s_1s_2Mm |\hat S_\nu |s_1s_2M'p\rangle \,
 \psi _{n}^{s_{1}s_{2}M'}(p).
\eeqa
Therefore, because the state $ |SM\rangle$
is an eigenstate of $\hat S_0$ and $\hat S_+\hat S_-$, the function
$\psi _{n}^{s_{1}s_{2}M}$ should likewise be an eigenfunction of
$\hat{\mathcal{S}}_0$ and $\hat{\mathcal{S}}_+\hat{\mathcal{S}}_-$ with the same eigenvalues.  With the expansions of Eqs.\  (\ref{eq:Mm}) and
(\ref{eq:sumS}), we obtain
\beqa
\hat{\mathcal{S}}_+\hat{\mathcal{S}}_- \psi _{n}^{s_{1}s_{2}M}(m)
&=& f_0(m) \psi _{n}^{s_{1}s_{2}M}(m) \nonumber \\
&&+ f_1(m)  \psi _{n}^{s_{1}s_{2}M}(m+1)
+ f_{-1}(m)  \psi _{n}^{s_{1}s_{2}M}(m-1), \label{eq:S+S-psi}
\eeqa
where
\beqa
f_0(m) &=& \langle s_1s_2Mm | \hat S_+  \hat S_-|s_1s_2Mm\rangle  ,\nonumber\\
f_1(m) &=& \langle s_1s_2Mm | \hat S_+  \hat S_-|s_1s_2  M,m+1\rangle  ,\nonumber\\
f_{-1}(m) &=& \langle s_1s_2Mm | \hat S_+  \hat S_-| s_1s_2 M,m-1\rangle = f_1(m-1) .
\eeqa

Equation (\ref{eq:S+S-psi}) can now be expressed in terms of finite difference operators, defined by
\beqa
\hat\Delta \psi(m) \equiv \hf\big( \psi(m+1)-\psi(m-1) \big),  \label{eq:FDiff1}\\
\hat\Delta^2\psi(m) \equiv \psi(m+1)- 2\psi(m) + \psi(m-1) ,  \label{eq:FDiff2}
\eeqa
with the result that
\beqa
f_1(m)\psi(m+ 1) =
f_1(m)\left(1  + \Delta + \hf \hat \Delta^2 \right) \psi(m), \\
f_1(m-1) \psi(m- 1) =
\left(1  - \Delta + \hf \hat \Delta^2 \right) {\left[f_1(m)\psi(m)\right]},
\eeqa
Eqn.\ (\ref{eq:S+S-psi}) then becomes
\beqa
\hat{\mathcal{S}}_+\hat{\mathcal{S}}_- \psi _{n}^{s_{1}s_{2}M}(m)
&=&  f_0(m)\psi _{n}^{s_{1}s_{2}M}(m)
+  f_1(m) \big[ 1+\hat \Delta + \hf \hat \Delta^2 \big] \psi _{n}^{s_{1}s_{2}M}(m)
\nonumber \\
&& +  \big[ 1-\hat \Delta + \hf \hat \Delta^2 \big] {\left[ f_1(m)\psi _{n}^{s_{1}s_{2}M}(m)
\right]},
\eeqa
and gathering terms leads to the expression of
$\hat{\mathcal{S}}_+\hat{\mathcal{S}}_-$ as the difference operator
\beq 
\hat{\mathcal{S}}_+\hat{\mathcal{S}}_- = F(m) + \hat \Delta f_1(m) \hat \Delta ,
\eeq
where
\beq
F(m) = f_0(m) + f_1(m) + f_1(m-1) .\label{bigFm}
\eeq
With $\hat S_+\hat S_-  = (\hat S^1_++\hat S^2_+)(\hat S^1_-+\hat S^2_-)$,
we also determine that
\beqa
f_0(m) &=&  \sigma_1^2- (M-m)(M-m-1) + \sigma_2^2-m(m-1), \label{f0m} \\
f_1(m) &=& \sqrt{ \big[\sigma_1^2 -(M-m)(M-m-1) \big]
\big[ \sigma_2^2 -m(m+1) \big] }, \label{f1m}
\eeqa
where
\beq
\sigma_i^2 = s_i(s_i+1), \quad i=1,2 .
\eeq

To determine asymptotic expressions for the functions, $\psi _{n}^{s_{1}s_{2}N}$, as eigenfunctions of
$\hat{\mathcal{S}}_+\hat{\mathcal{S}}_-$, we now make the \emph{continuous variable approximation}
 of extending these functions of the discrete variable $m$ to functions, $\Psi _{n}^{s_{1}s_{2}N}$, of a continuous variable $x$ with the property that
\beq
\Psi_n^{s_{1}s_{2}N}(x) = \psi _{n}^{s_{1}s_{2}N}(x),
\eeq
 whenever $x$ is in the domain of the discrete variable $m$.
In this approximation, which is valid in the asymptotic limits in which
$\Psi_n^{s_{1}s_{2}N}$ becomes a smooth function,
the difference operators can be replaced by differential operators:
 \beq
 \hat \Delta\to \hat D\equiv \frac{d}{dx}\, ,\quad \hat \Delta^2\to \hat D^2\equiv \frac{d^2}{dx^2}\, .
 \eeq
 The functions $F$, $f_0$ and $f_1$ of Eqns.\ (\ref{bigFm})--(\ref{f1m}) are similarly extended to the continuous variable $x$, and the operator
 $\hat{\mathcal{S}}_+\hat{\mathcal{S}}_-$ becomes the differential operator
\beq
\hat{\mathcal{S}}_+\hat{\mathcal{S}}_- \to F(x) + \hat D f_1(x) \hat D .
\eeq

Provided the extension of  $ \psi _{n}^{s_{1}s_{2}M}(m)$ to the function $ \Psi _{n}^{s_{1}s_{2}M}(x)$ does not require the latter to be non-zero for any $x$ that is outside of the limits for $m$, it is seen that $f_1(x)$ is real for all $x$ for which $ \Psi _{n}^{s_{1}s_{2}M}(x)$ is non-zero.
The limits on the values of $m$ are seen, from Eqn.\ (\ref{eq:Mm}), to be such that
$-s_2 \leq m\leq s_2$ and $-s_1 \leq M-m \leq s_1$.  We will denote the upper and lower limits on the value of $m$, by $m_{\scriptstyle max }$ and $m_{\scriptstyle min }$, respectively.
 Then, because the norm of the function    $ \psi _{n}^{s_{1}s_{2}M}$ is given by
\beq
\|\psi _{n}^{s_{1}s_{2}M}\|^2 =
\sum_{m = m_{\scriptscriptstyle min}}^{m_{\scriptscriptstyle max}} |\psi _{n}^{s_{1}s_{2}M}(m)|^2 ,
\eeq
it follows that the corresponding
smooth function $ \Psi _{n}^{s_{1}s_{2}M}(x)$ should have norm given by
\beq
\|\Psi _{n}^{s_{1}s_{2}M}\|^2 = \int_{m_{\scriptscriptstyle min}}^{m_{\scriptscriptstyle max}} |\Psi _{n}^{s_{1}s_{2}M}(x)|^2\, dx .
\eeq
It is also seen that, when $\Psi _{n}^{s_1s_2M}(x)$ is zero for all
$x>m_{\scriptstyle max} $ and all $x<m_{\scriptstyle min }$, this integral can be extended to the range $-\infty < x<\infty$.  The operator $\hat D=d/dx$ is then seen to be skew Hermitian and
$\hat{\mathcal{S}}_+\hat{\mathcal{S}}_-$ is Hermitian.

Now, if the function $\Psi _{n}^{s_{1}s_{2}M}$ is sufficiently smooth, is  non-zero over a sufficiently narrow region of $x$ within the limits
$m_{\scriptstyle min} < x < m_{\scriptstyle max}$ and is  centered about a value $x_0$, we can make the so-called SHA \cite{Chen,SHAIBM} which,
in addition to the continuous variable approximation,
 consists of dropping all  terms in $F, f_0$ and $f_1$  
  for which the expansion of the operator $\hat S_+ \hat S_-$  will be  more than bilinear in $x - x_0$ and $d/dx$.
The conditions under which the SHA are valid are shown, in the following,
to be well satisfied, for finite values of $n$, in the asymptotic limit as $s_1\to \infty$ and $s_2\to\infty$.
Thus, the SHA gives the asymptotic expression
\beq
\hat{\mathcal{S}}_+\hat{\mathcal{S}}_- \approx E  + \half A \frac{d^2}{dx^2}
+C (x-x_0) - \half B (x-x_0)^2 + \dots ,
\eeq
where
\beq
E = F(x_0) , \quad A= 2 f_1(x_0) ,\quad
C= F'(x_0) ,\quad B= -F^{\prime\prime}(x_0) .        \label{EACB}
\eeq

An examination of the values of these parameters reveals that
$A>0$ and $B>0$  in situations of interest.
Thus, we consider
${\cal H} \approx - \hat{\mathcal{S}}_+\hat{\mathcal{S}}_- $
and determine $x_0$ to be the value for which  $C=0$, bringing ${\cal H}$ to the form
\beq
{\cal H} = -E  +
\left[-\frac{1}{2a^2} \frac{d^2}{dx^2} + \frac{1}{2}a^2 (x-x_0)^2\right] \hbar\omega ,
\eeq
where
\beq
(\hbar\omega)^2 = AB, \quad a^4 = \frac{B}{A} .
\eeq
The eigenfunctions for this Hamiltonian are  harmonic oscillator wave functions, and the eigenvalues are
\beq E_n  = -E + \big( n+\hf \big)\hbar\omega. \eeq
Thus, we obtain  the desired asymptotic Clebsch-Gordan coefficients in the form given by Eqn.\ (\ref{mainsu2result}).
Note that the eigenfunctions of a Hamiltonian are only ever defined to within an arbitrary phase.  There are also arbitrary phases in the definition of CG coefficients. The phases in Eqn.\ (\ref{mainsu2result}) were chosen to reproduce the standard phases of the Condon and Shortley CG coefficients \cite{ConShort}.

\subsection{Simplified analytical expressions for asymptotic SU(2){CG } coefficients} \label{sect:limitsu2}

The SHA results given above are easily calculated, and give accurate results in the limit as $s_1\to \infty$ and $s_2\to\infty$ but $M$ and $n$ remain finite.
Simpler analytic asymptotic expressions are obtained if we also  neglect terms that go to zero as $s_1\to \infty$ and $s_2\to\infty$.

In these limits
\beq {\cal H} \to {\rm const.} - \half A_0 \frac{d^2}{dx^2}
+C_0(x-x_0) +  \half B_0 (x-x_0)^2  ,
\eeq
with
\beqa
A_0 &\sim& 2\sigma_1\sigma_2
\left( 1-\frac{M^2}{(\sigma_1+\sigma_2)^2}\right),
\\
C_0 &\sim& \frac{2(\sigma_1+\sigma_2)}{\sigma_1}M , \\
B_0&\sim& \frac{2(\sigma_1+\sigma_2)^2}{\sigma_1\sigma_2}  .
\eeqa
Thus, we can evaluate the asymptotic Clebsch-Gordan coefficients from
Eqn.~(\ref{mainsu2result}) with
\beqa 
a^4 &=& \frac{B_0}{A_0} =\frac{(\sigma_1+\sigma_2)^4}
{\sigma_1^2\sigma_2^2 \big[ (\sigma_1+\sigma_2)^2-M^2\big]}\, , \label{eq:49}\\
x_0&=&\frac{C_0}{B_0} = \frac{\sigma_2 M}{\sigma_1+\sigma_2},  \label{eq:50}\eeqa
to obtain the expression
\beqa
&&( s_{1},\!M\!-\!m,\; s_{2}m |s_{1}\!+\!s_{2}\!-n,M)\nonumber \\
&&\qquad \sim
(-1)^n \Big(\frac{a}{\sqrt{\pi} 2^n n!}\Big)^{\frac12}
H_n\big( a (m-x_0)\big) e^{-\frac12 a^2(m-x_0)^2} , \label{eq:su2_asym}
\eeqa
and that of Eq. (\ref{mainsu2result2}).

\subsection{Numerical  results for su(2) CG coefficients}

We have ascertained and the following examples illustrate that the SU(2) CG coefficients satisfy the conditions for the validity of the SHA in the specified asymptotic limit as $s_1$, $s_2\to\infty$ for finite values of $n$ and $M$.
The following results show that in addition to being precise in these limits, the SHA and simplified SHA wave functions also give remarkably accurate values for modest values of $s_1$ and $s_2$ and surprising large values of $M$.

The following figures compare  the values of exactly computed SU(2) CG coefficients,
$(s_{1}, M\!-\!m ,\, s_{2}\,m |s_{1}\!+\!s_{2}\!-n,M)$, with the SHA expressions given by Eqn.~(\ref{mainsu2result}).  The continuous (red) lines are those of the full SHA approximation with the parameters as defined in Sect.~\ref{sect:su2SHA}.  The dashed (black) lines are those of the simplified SHA coefficients with the parameters given by Eqns.~(\ref{eq:49} -- \ref{eq:50}).

\begin{figure}[hpt]
\begin{center}
\epsfig{file=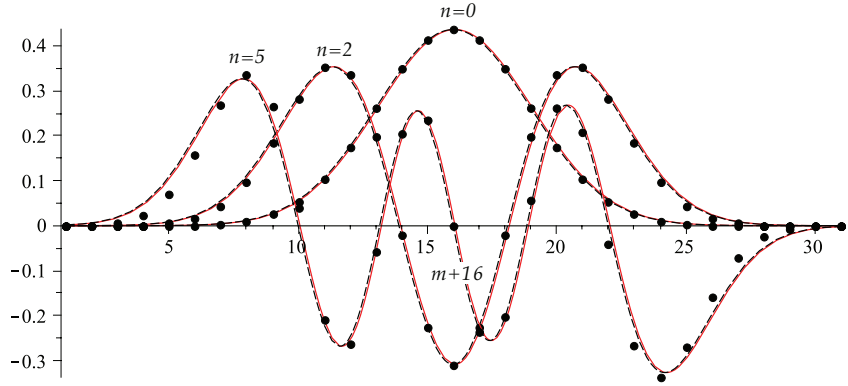, height=2in}
\caption{The Clebsch-Gordan coefficients
$(20,-m, 15,m|35-n,0)$ shown as a function of $m$ for three values of $n$.  Exact values are shown as
dots, full SHA values as continuous (red) lines, and simplified SHA values as dashed lines.
\label{fig:a}}
\end{center}
\end{figure}

Figure \ref{fig:a}  illustrates that, for $s_1$ and $s_2$ as small as 20 and 15, respectively, both the full and simplified SHA yield approximate SU(2) CG coefficients for $n\leq 5$ that are almost indistinguishable from each other. The figure shows that inaccuracies become visible for $n=5$ coefficients when $|m| \gtrsim 8$.  The results
become increasingly accurate for larger values of $s_1$ and $s_2$ and are precise in the asymptotic
limit.

\begin{figure}[hpt]
\begin{center}
\epsfig{file=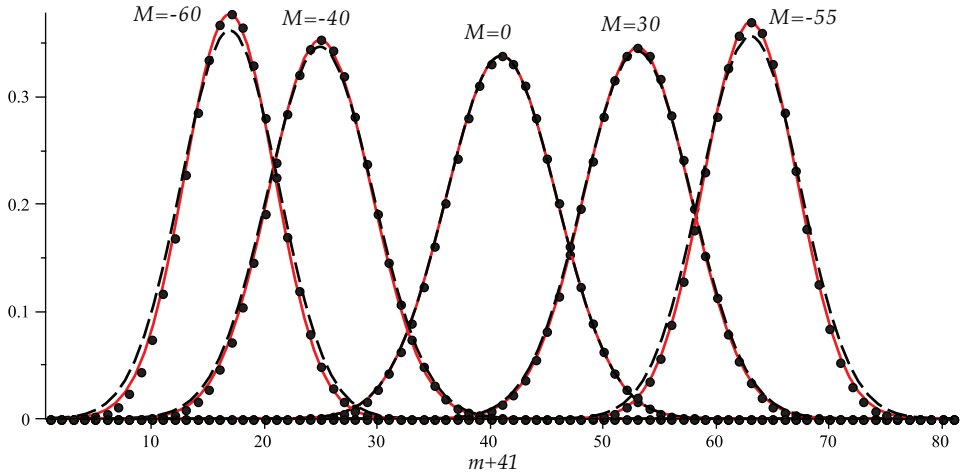, height=2.5in}
\caption{The Clebsch-Gordan coefficients
$(60,M-m, 40,m|100,M)$ shown as a function of $m$ for a range of $M$ values.   Exact values are shown as
dots, full SHA values as continuous (red) lines, and simplified SHA values as dashed lines. \label{fig:b}}
\end{center}
\end{figure}

Figure \ref{fig:b} illustrates  that, for $s_1=60$, $s_2=40$, and $n=0$ the simplified SHA is very accurate for $|M| \lesssim 30$ whereas the full SHA is accurate for all the values  of $M$ shown.

\begin{figure}[hpt]
\begin{center}
\epsfig{file=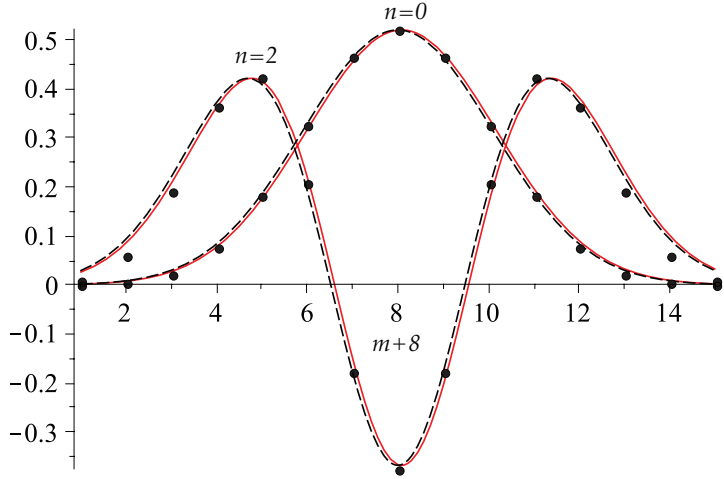, height=2.5in}
\caption{The Clebsch-Gordan coefficients
$(10,-m, 7,m|17-n,0)$ shown as a function of $m$.   Exact values are shown as
full dots, full SHA values as continuous (red) lines, and simplified SHA values as dashed lines. \label{fig:c}}
\end{center}
\end{figure}

Figure \ref{fig:c} illustrates how  accurate the SHA and simplified SHA Clebsch-Gordan coefficients can be for quite small values of $s_1$ and $s_2$ provided $n$ and $|M|$ are kept even smaller.
The region in which the results are at their worst is for values of $m$ close to its boundary values, especially in situations in which the asymptotic expressions extend beyond these boundaries.

\section{Asymptotic SU(1,1) Clebsch-Gordan coefficients}\label{sec:su11theory}

We now consider the operators $\{\hat{K}_{+},\hat{K}_{-},\hat{K}_{0}\}$, that satisfy the commutation relations of the complex extension of the su(1,1) Lie algebra:
\begin{equation}
\lbrack \hat{K}_{0},\hat{K}_{\pm }]=\pm \hat{K}_{\pm },\quad
\lbrack \hat{K}_{-},\hat{K}_{+}]=2\hat K_{0}.  \label{su11}
\end{equation}
Basis states, $\{ |kn\rangle, n=0,1,2,\dots\}$, for a unitary su(1,1) irrep of the positive discrete series are defined by the equations
\beqa
\hat K_0 |k  n\rangle =(k+n) |k n\rangle, \nonumber \\
 \hat K_+ |k  n\rangle = \sqrt{(2k +n)(n+1)}\, \vert k,n+1\rangle, \nonumber  \\
 \hat K_- |k  n\rangle =\sqrt{(2k +n-1)n}\, |k ,n-1\rangle ,  \label{eq:1.su11S0+-} \\
 \hat K_+\hat K_- |k n\rangle = (2k + n-1)n |k n\rangle . \nonumber
\eeqa

Basis states for irreps of su(1,1) within the tensor product of two such irreps are  given by linear combinations
\beq
\vert Kn\rangle = \sum_{n_1n_2} |k_2n_2\rangle \otimes |k_1n_1\rangle\,
C(k_1,n_1, k_2,n_2; K,n),
\eeq
where $C(k_1,n_1, k_2,n_2; K,n) $  is an SU(1,1) CG coefficient in the notation of Van der Jeugt \cite{VdJ},
equal to the overlap  of a coupled and uncoupled tensor product states
\beq
C(k_1,n_1, k_2,n_2; K,n)
= \big[ \langle k_1,n_1|\otimes \langle k_2,n_2| \big] |Kn\rangle  .
 \eeq
Thus, if we put  $n_1 = \hf N-m$ and  $n_2 = \hf N+m$ and denote an uncoupled  tensor product state by
\beq
|k_1k_2Nm\rangle \equiv  |k_2, \hf N +m\rangle \otimes |k_1,\hf N-m\rangle, \label{eq:Nm}
\eeq
we obtain the more useful expression of a CG coefficient as an overlap
\beq
C(k_1,n_1, k_2,n_2; K,n) 
=  \langle k_1 k_2 Nm |Kn\rangle  .
\eeq

These coefficients are defined (to within arbitrary phase factors) by the requirement that the states $\{ |Kn\rangle\}$ satisfy the eigenvalue
 equations
\beqa
\hat K_0 |Kn\rangle = (K+n)|Kn\rangle, \label{eq:S0Kn}\\
    \hat K_+\hat K_-  |Kn\rangle = (2K+n-1)n |Kn\rangle,
\eeqa
with
\beq
\hat K_0 = \hat K_0^1 + \hat K_0^2 , \quad
\hat K_\pm = \hat K_\pm^1 + \hat K_\pm^2 , \label{eq:sumS2}
\eeq
and the understanding that
\beqa
\hat K^1_\nu\big(  |k_2 m_2\rangle \otimes |k_1m_1\rangle \big)
&=&   |k_2 m_2\rangle \otimes \hat K^1_\nu |k_1m_1\rangle  ,
\\
\hat K^2_\nu\big(  |k_2 m_2\rangle \otimes |k_1m_1\rangle \big)
&=& \hat K^2_\nu  |k_2 m_2\rangle \otimes |k_1m_1\rangle .
\eeqa

\subsection{SU(1,1) CG coefficients in the shifted harmonic approximation}
\label{sect:su11SHA}

We now determine asymptotic limits to these CG coefficients in the SHA and show them to be precise for large values of $N$ and finite values of $n$ and
$|k_1-k_2|$.

Before embarking on an SHA calculation of asymptotic SU(1,1) CG coefficients, we first examine some coefficients to see if they satisfy the necessary conditions for the validity of the SHA.
From exact calculations in the phase convention of Van der Jeugt
\cite{su11code}, it is determined that, for large values of $N$ and small values of $n$, the SU(1,1) CG coefficients $\langle k_1 k_2Nm \vert Kn\rangle$,
when multiplied by the phase factor $(-1)^{N+m}$ and
 regarded as functions of $m$, approach standard harmonic oscillator wave functions.
 Thus, we define an overlap function of the discrete variable $m$ by
\beq
\psi^{k_1k_2N}_n(m) = (-1)^{N+m} \langle k_1k_2Nm |Kn\rangle
\eeq
with  $K = N-n + k_1 + k_2$.

Because the state $|Kn\rangle$ is an eigenstate of the operator
$\hat K_+\hat K_-$, it follows that the representation of this state by the function $ \psi^{k_1k_2N}_n$, of the discrete variable $m$ is an eigenfunction of the operator $\hat{\cal K}_+\hat{\cal K}_-$ defined by
\beqa
\hat{\cal{K}}_+\hat{\cal K}_- \psi _{n}^{k_{1}k_{2}N}(m)
&\equiv& (-1)^{N+m} \langle k_1k_2Nm |\hat K_+\hat K_- |Kn\rangle \nonumber \\
&=&\sum_p (-1)^{m-p}  \langle k_1k_2Nm |\hat K_+\hat K_- |k_1k_2 Np\rangle
\, \psi _{n}^{k_{1}k_{2}N}(p).
\eeqa
Thus, we obtain
an equation for $\psi _{n}^{k_{1}k_{2}N}$ of identical form to that of Eqn.\ (\ref{eq:S+S-psi}) for the SU(2) CG coefficients, given here by
\beqa
\hat{\mathcal{K}}_+\hat{\mathcal{K}}_- \psi _{n}^{k_{1}k_{2}N}(m)
&=&  f_0(m) \psi _{n}^{k_{1}k_{2}N}(m)
+ f_1(m)  \psi _{n}^{k_{1}k_{2}N}(m+1) \nonumber \\
&&+ f_{1}(m-1)  \psi _{n}^{k_{1}k_{2}N}(m-1), \label{eq:S+S-2}
\eeqa
but now with
\beqa
f_0(m) &=& \langle k_1k_2 Nm | \hat K_+  \hat K_-|  k_1k_2 Nm\rangle  \nonumber\\
&=&  (2k_1 +\hf N-m -1) (\hf N-m)  \nonumber \\
&&\qquad  +  \big(2k_2 +\hf N+m -1\big) \big(\hf N+m \big) , \\
f_1(m) &=&- \langle  k_1k_2Nm | \hat K_+  \hat K_-|  k_1k_2N,m+1\rangle  \nonumber\\
&=& - \left[ \big(2k_1 +\hf N-m-1\big) \big(\hf N-m \big)\right.
\nonumber \\
&&\qquad\left. \times      \big(2k_2 +\hf N+m\big) \big(\hf N+m+1 \big)\right]^{\frac12} .
\eeqa
To determine asymptotic expressions for the functions,
$\psi _{n}^{k_{1}k_{2}N}(m)$, as eigenfunctions of
$\hat{\mathcal{K}}_+\hat{\mathcal{K}}_-$, we
extend these functions of the discrete variable $m$ to functions,
$\Psi _{n}^{k_{1}k_{2}N}$, of a continuous variable $x$ with the property that
\beq
\Psi_n^{k_{1}k_{2}N}(x) = \psi _{n}^{k_{1}k_{2}N}(x),
\eeq
whenever $x$ is  in the domain of the discrete variable $m$.
Thus, as for SU(2), we obtain an expression for
$\hat{\cal{K}}_+\hat{\cal K}_- $ as the differential operator
\beq
\hat{\cal{K}}_+\hat{\cal K}_- = F(x) + \hat D f_1(x) \hat D ,
\eeq
where  $F(x) = f_0(x) + f_1(x) + f_1(x-1)$ and $\hat D = d/dx$.
With $\kappa_1$ and $\kappa_2$ defined by
\beq
2\kappa_1 := 2k_1+\hf N , \quad 2\kappa_2 := 2k_2+\hf N,
\eeq
we also obtain the expressions
\beqa
f_0(x) &=&  \big(2\kappa_1 -x -1\big) \big(\hf N -x \big)
            +  \big(2\kappa_2 +x -1\big) \big(\hf N +x \big)   \\
f_1(x) &=& - \big[ \big(2\kappa_1 -x-1\big) \big(\hf N -x \big)
            \big(2\kappa_2  +x\big) \big(\hf N +x+1 \big)\big]^{\frac12} .
\eeqa

Provided the extension of  $ \psi _{n}^{k_{1}k_{2}N}(m)$ to the smooth function $ \Psi _{n}^{k_{1}k_{2}N}(x)$ does not require the latter to be non-zero for any $x$ that is outside of the limits for $m$, it is seen that $f_1(x)$ is real for all $x$ for which $ \Psi _{n}^{s_{1}s_{2}M}(x)$ is non-zero.
The limits on the values of $m$ are seen, from Eqn.\ (\ref{eq:Nm}), to be such that
$-N/2 \leq m\leq N/2$.  Then, because the norm of the function
$ \psi _{n}^{k_{1}k_{2}N}(m)$ is given by
\beq
\| \psi _{n}^{k_{1}k_{2}N}(m)\|^2 =
\sum_{m = -N/2}^{N/2} |\ \psi _{n}^{k_{1}k_{2}N}(m)|^2 ,
\eeq
it follows that the corresponding
smooth function $\Psi _{n}^{k_{1}k_{2}N}(x)$ should have norm given by
\beq
\|\Psi _{n}^{k_{1}k_{2}N}\|^2 =
 \int_{-N/2}^{N/2} |\Psi _{n}^{k_{1}k_{2}N}(x)|^2\, dx .
 \eeq
Also, when evaluated without approximation,
$\Psi _{n}^{k_{1}k_{2}N}(x)$ is zero for all $x>N/2$ and all $x<-N/2$ and this integral can  be extended to the range $-\infty < x \red{<} \infty$.  The operator $\hat D=d/dx$ is then seen to be skew Hermitian and
$\hat{\mathcal{K}}_+\hat{\mathcal{K}}_-$ is Hermitian.

Now, if the function $\Psi _{n}^{k_{1}k_{2}N}(x)$ is sufficiently smooth, is non-zero over a narrow region of $x$ within the limits
$-N/2 < x < N/2$, and is centered about a value $x_0$,
we can again invoke the SHA of dropping all  terms that are more than bilinear in $x-x_0$
and $d/dx$ in an expansion of the operator
$\hat{\mathcal{K}}_+\hat{\mathcal{K}}_-$.
This gives
\beq \hat{\mathcal{K}}_+\hat{\mathcal{K}}_- \approx E  - \half A \frac{d^2}{dx^2}
+C (x-x_0) + \half B (x-x_0)^2  , \eeq
where
\beq E = F(x_0) , \quad  A= -2 f_1(x_0) , \quad
C= F'(x_0) , \quad B=  F^{\prime\prime}(x_0) .
\eeq
This approximation becomes precise,
in the $N\to \infty$ asymptotic limit, provided the parameters $n$, $k_1$, and $k_2$ remain finite.  In fact, it turns out that $k_1+k_2$ can also be large provided the difference $|k_1-k_2|$ remains small in comparison to $N$.  The manner in which the SHA ceases to be valid, for large values of $n$ and $|k_1-k_2|$, relative to $N$, is shown in Sect.\ \ref{sect:su11results}.
Thus, we consider the SHA Hamiltonian
\beq {\cal H} = E  - \half A \frac{d^2}{dx^2}
+C (x-x_0) +  \half B (x-x_0)^2  . \eeq

The appropriate value for $x_0$ is that
for which $C=0$ and
\beq {\cal H} = E  +
\left[-\frac{1}{2a^2} \frac{d^2}{dx^2} + \frac{1}{2}a^2 (x-x_0)^2\right] \hbar\omega ,
\eeq
with
\beq
(\hbar\omega)^2 = AB, \quad a^4 = \frac{B}{A} .
\eeq
The eigenfunctions for this Hamiltonian are  harmonic oscillator wave functions,
and the eigenvalues are given by
\beq
E_n  = E + \big( n+\hf \big)\hbar\omega.
\eeq
With a standard choice of phase for the harmonic oscillator eigenfunctions, we then obtain the desired asymptotic Clebsch-Gordan coefficients, as given in Eqn.\ (\ref{eq:ASSU11coefs}).

\subsection{Simplified analytical expressions for asymptotic SU(1,1) {CG } coefficients}

The SHA results given above are easily calculated, and give accurate results  for large values of $N\to \infty$ and small values of $n$.
Simpler analytical expressions are obtained if we further neglect terms in the SHA Hamiltonian which go to zero in the $N\to \infty$ asymptotic limit.
In these limits
\beq {\cal H} \to {\rm const.}  - \half A_0 \frac{d^2}{dx^2}
+C_0x +  \half B x^2  , \eeq
with
\beqa
A_0 &=& 2N\sqrt{\kappa_1\kappa_2},\\
C_0 &=&
 \frac{N+ 4\sqrt{\kappa_1\kappa_2}}{2\sqrt{\kappa_1\kappa_2}}
 (\kappa_2-\kappa_1), \\
 B_0 &=&
\frac{\big(N+ 4\sqrt{\kappa_1\kappa_2}\big)^2}{2N\sqrt{\kappa_1\kappa_2} }.
\eeqa
Thus, we can evaluate the asymptotic Clebsch-Gordan coefficients from
Eqn.~(\ref{eq:ASSU11coefs}) with $a^2=\sqrt{B_0/A_0}$ and $x_0=-C_0/B_0$ given explicitly
in Eqn.~(\ref{eq:su11analytic})

\subsection{Numerical  results for su(1,1) CG coefficients} \label{sect:su11results}

We have ascertained and the following examples illustrate that the SU(1,1) CG coefficients, when multiplied by a phase factor
$(-1)^{N+m}$, satisfy the conditions for the validity of the SHA in the specified asymptotic limit as $N\to\infty$ for finite values of $n$ and $|k_1-k_2|$.
The restriction on the value of $n$ is because the harmonic oscillator wave functions become broad and oscillate rapidly for large values of $n$ with the result that if $n/N$ is too large the conditions for the validity of the SHA cease to be satisfied.  Likewise,
the restriction on the value of $|k_1-k_2|$ is understood to arise from the value of $C_0$ which, for $N\to\infty$, approaches $C_0\sim 3(k_2-k_1)$.
Thus, unless $|k_1-k_2| << N$, it is not guaranteed that the centroid, $x_0\approx -C_0/B_0$, of the SHA wave function will be sufficiently close to the center of the domain $-N/2 \leq x \leq N/2$ for the SHA to be a valid approximation.
However, the  results below show that in addition to being precise in the asymptotic limits, the SHA (as opposed to the simplified SHA) remains remarkably accurate for
relatively large values of $|k_1-k_2|$ when $N$ is large.

The following figures compare  the values of exactly computed SU(1,1) CG coefficients,
$\langle k_1k_2 Nm|K n\rangle$, with the SHA expressions given by Eqn.~(\ref{eq:ASSU11coefs}).  The continuous (red) lines are those of the full SHA approximation with the parameters as defined in sect.~\ref{sect:su11SHA}.  The (black) dashed  lines are those of the simplified analytical   SHA coefficients with parameters given by Eqn.~(\ref{eq:su11analytic}). The asymptotic limits require that $n$ remains finite while $N\to\infty$.
The simplified SHA coefficients retain the property of being precise in the $N\to \infty$ limit but requires, in addition, that
$|k_1-k_2|$ remains small compared to $N/2$.  In fact, as the following figures illustrate the SHA coefficients yield surprisingly good approximations for quite modest values of $N$ and the analytical approximations are seen to be almost indistinguishable from the full SHA expressions when $|k_1-k_2|$ is not too large.

\begin{figure}[hpt]
\begin{center}
\epsfig{file=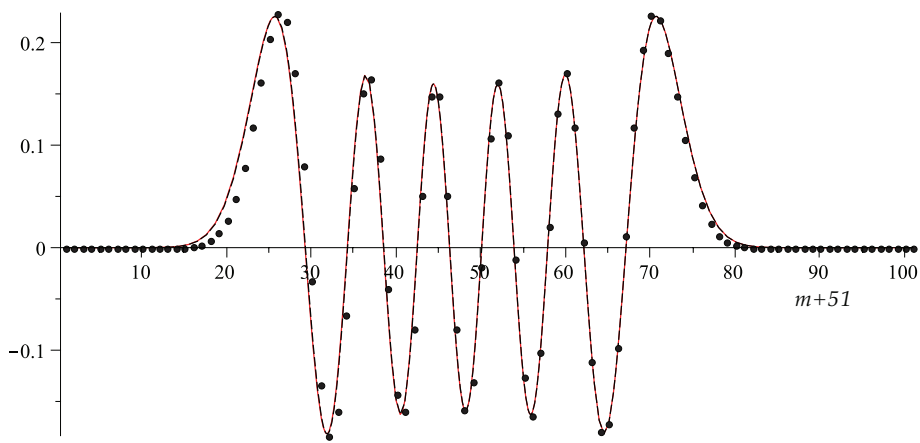, height=2.25in}
\caption{SU(1,1) Clebsch-Gordan coefficients
$(-1)^m \langle 10, 17, 100,m | 117,10\rangle$ shown as a function of
$m$.  Exact values are shown as full dots, full SHA values as continuous (red) lines, and simplified SHA values as dashed lines.
\label{fig:d}}
\end{center}
\end{figure}

Figure \ref{fig:d}  illustrates
the accuracy that can be obtained with the above-defined asymptotic SU(1,1) CG coefficients for $N=100$, $n=10$ and $k_2-k_1 = 7$.
It shows that
the SHA coefficients and the analytical approximations to them are virtually indistinguishable.  It also shows that the errors in the asymptotic coefficients for a large but finite value of $N$ start to become most noticeable,
for these relatively large values of $n$ and $|k_1-k_2|$,
 at the upper and lower reaches of $m$.

\begin{figure}[hpt]
\begin{center}
\epsfig{file=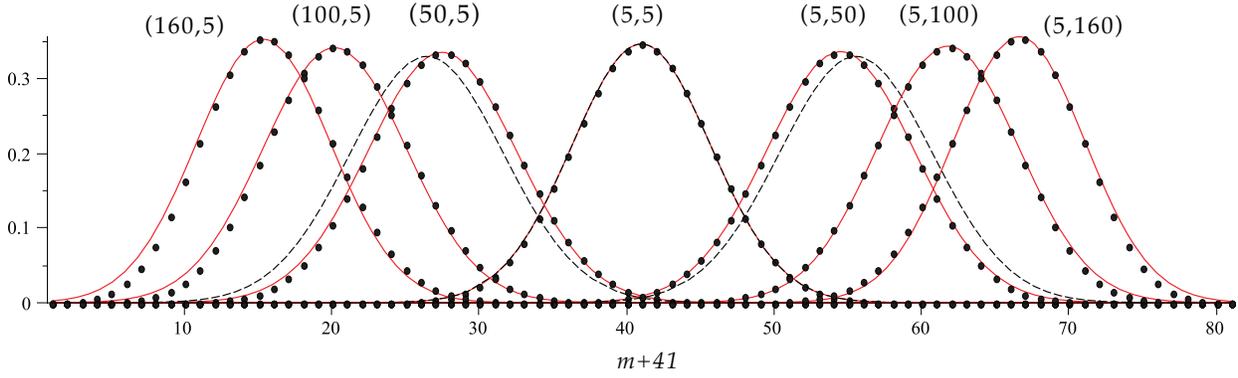, height=2.in}
\caption{SU(1,1) Clebsch-Gordan coefficients
$(-1)^m \langle k_1 k_2,80,m| K,0\rangle$ with $K = 80 + k_1+k_2$
for a range of $(k_1,k_2)$ values shown as  functions of $m$.   Exact values are shown as
full dots, full SHA values as continuous (red) lines, and simplified SHA values as dashed lines.
The simplified SHA values for $|k_1-k_2| > 45$, are unacceptably inaccurate and are not shown.
\label{fig:e}}
\end{center}
\end{figure}

Figure \ref{fig:e} shows the SU(1,1) CG coefficients
for a range of $k_1-k_2$ values.  The calculations for other  $k_1-k_2$ values  show that the simplified SHA coefficients are
close to those of the full SHA for $|k_1-k_2| \lesssim 25$ but are noticeably different for larger values of
$|k_1-k_2|$ as seen for the $(k_1,k_2) = (5,50)$ and (50,5) coefficients shown in the figure.
It is also seen that, for the full SHA, errors start to become evident for large values of $|k_1-k_2|$ as the value of $m$ approaches its upper or lower bounds.

\begin{figure}[hpt]
\begin{center}
\epsfig{file=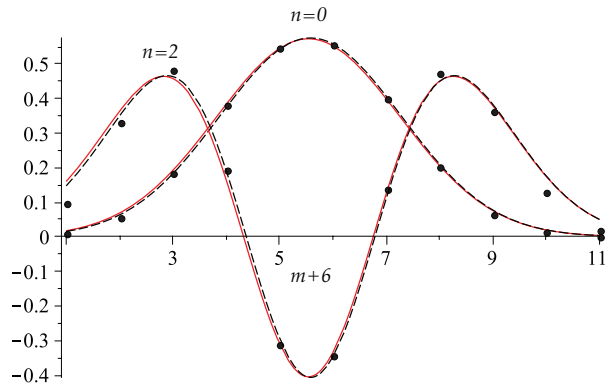, height=2.in}
\caption{SU(1,1) Clebsch-Gordan coefficients
$(-1)^m\langle 1/2,3/2, 10,m |12,0\rangle$  and
$\langle 1/2,3/2,10,m |10,2\rangle$  shown as a function of $m$.   Exact values are shown as
full dots, full SHA values as continuous (red) lines, and analytical asymptotic limits of the SHA values as dashed lines.
\label{fig:f}}
\end{center}
\end{figure}

The  SU(1,1) CG coefficients, $\langle k_1k_2 Nm | Kn\rangle $,
 are given exactly
both in the SHA and in the simplified analytical approximation to the SHA,
for finite values of $n$ and $|k_1-k_2|$,  in the $N\to\infty$ asymptotic limit.
However, the full SHA expressions remain noticeably more accurate over a considerably larger domain.
In fact,  the coefficients continue to be accurate
for relatively small values of $N$ and even smaller values of $n$ and $|k_1-k_2|$.
This is illustrated for $N=10$ in fig.~\ref{fig:f}.
The region in which the results are at their worst is for values of $m$ close to their limits, especially in situations in which the asymptotic expressions extend beyond these limits.

\section{Comparisons with the random-phase approximation (RPA)}

Whereas the SHA and RPA are both harmonic oscillator approximations and both become precise in asymptotic limits, they apply in different complementary situations.  In their application to the derivation of asymptotic SU(2) and SU(1,1) CG coeficients, the RPA can be viewed as a contraction of the su(2) Lie algebra to a harmonic oscillator boson algebra.  Thus, we consider the possibility that the SHA might also correspond to such a contraction, albeit one that is valid in a different domain of an SU(2) representation space.

\subsection{SU(2) {CG } coefficients}

We start from the observation that SU(2) CG coefficients are the eigenstates of a
Hamiltonian
\beq \hat H = \alpha\hat S_0  + \chi \hat S_+ \hat S_-
\label{Happrox}\eeq
on the tensor product space of two su(2) irreps,
$\{s_1\}\otimes \{s_2\}$ in a basis of product states
$\{ |s_1m_1\rangle \otimes |s_2m_2\rangle\}$.
 The Hamiltonian $\hat H$ has eigenstates, $\{ |SM\rangle\}$, with eigenvalues given by
\beq E_{SM} =  \alpha M + \chi \big[ S(S+1) - M(M-1)\big]  .
\eeq
Inspection shows that, when $\chi>-\alpha$, for $\alpha >0$, the lowest-energy eigenstate of $\hat H$ is the state with $S=\sigma$, $M=-S$, where
$\sigma =s_1+s_2$ and, when $\chi < -\alpha$, the lowest-energy eigenstate is the state with $S=\sigma$, $M=0$.  Thus, the two situations are relevant for the calculation of asymptotic SU(2) CG coefficients $(s_1 m_1\, s_2m_2 |SM)$, for large values of  $s_1$, $s_2$, and $S$, with $M$ close to either $-S$ or 0, respectively.
As we now observe, RPA  gives solutions in the first scenario whereas the SHA does so in the second.

In the  $\sigma\to\infty$ asymptotic limit for $\alpha >0$ and $\chi>-\alpha$,
 it is convenient to relabel the low-lying states by $|SM\rangle \to |nm\rangle$, where $n=\sigma-S$ and $m= S+M$.  The energy
$\mathcal{E}_{nm} = E_{SM}$ is then given to leading order in $n$ and $m$ by
\beq \mathcal{E}_{nm} \sim -  \alpha \sigma +  \alpha n +
[ \alpha +\chi(2\sigma+1)] m , \eeq
which is an eigenvalue of a Hamiltonian expressed in terms of the raising and lowering operators of two simple harmonic oscillators by
\beq \hat \mathcal{H} = - \alpha\sigma +  \alpha c^\dag c +
[ \alpha+\chi(2\sigma+1)] b^\dag b, \label{eq:RPAHam}\eeq
with
\beq c^\dag c |nm\rangle = n |nm\rangle , \quad
b^\dag b |nm\rangle = m |nm\rangle .\eeq
For  asymptotically large values of $s_1$ and $s_2$, this result can be obtained in the RPA by contracting the su(2) algebra in each of the $\{ s_1\}$ and $\{ s_2\}$ irreps by
\beq \hat S_+ \sim \frac{1}{\sqrt{s}}\, a^\dag , \quad \hat S_- \sim \frac{1}{\sqrt{s}}\, a , \quad \hat S_0 = -s + a^\dag a , \label{eq:su2contraction}\eeq
for $s=s_1$ and $s=s_2$, respectively.
Diagonalization of  the  Hamiltonian $\hat H$ of Eqn.\ (\ref{Happrox}) in this contraction limit, then leads to the asymptotic expression (\ref{eq:RPAHam}) and determines the $c$ and $b$ boson operators from which one can derive asymptotic CG coefficients.
The contraction (\ref{eq:su2contraction}) which leads to this RPA result is valid in the domain of states, $\{ |SM\rangle\}$, in which $M$ is close to $-S$ and $S$ is close to $s_1+s_2$, for large values of $s_1$ and $s_2$.

For $\chi< -\alpha$, the RPA breaks down because the ground state of the Hamiltonian suddenly flips from an $|S,M=-S\rangle$ state to an
$|S,M=0\rangle$ state at $\chi=-\alpha$. However, for $\chi< -\alpha$ the SHA provides asymptotic solutions.

In the  $\sigma\to\infty$ asymptotic limit for $\chi< - \alpha$, it is convenient to relabel the low-lying states by
$|SM\rangle \to |nM\rangle$, where $n=\sigma-S$. It is then found that
$\mathcal{E}_{nM} = E_{SM}$ is given to leading order in $n$  by
\beq \mathcal{E}_{nM} \approx   \chi \sigma(\sigma+1)- \chi(2\sigma+1)n +
 \alpha M -  \chi M(M-1) . \eeq
Thus, in this asymptotic limit, the spectrum of eigenvalues of the Hamiltonian (\ref{Happrox}) is that of a simple harmonic oscillator coupled to a U(1) rotor
\beq \hat \mathcal{H} =   \chi \sigma(\sigma+1) - \chi(2\sigma+1)c^\dag c
+  \alpha \hat S_0 -  \chi \hat S_0(\hat S_0-1)    ,  \label{eq:su2_sha_limit}\eeq
where
\beq c^\dag c |nM\rangle = n |nM\rangle , \quad
\hat S_0 |nM\rangle = M |nM\rangle .\eeq

The SHA gives an explicit expression for a shifted harmonic oscillator that is essentially equivalent to $\hat\mathcal{H}$.  We have not succeeded in deriving this Hamiltonian by a contraction of the su(2) algebra.
However, it is noted that the SHA is based on a realisation of the su(2) Lie algebra given by the expressions, in terms of the harmonic oscillator operators,
$\hat x=x$, $\hat p= -{\rm i}\hbar d/dx$ \beqa \hat S_+ \to \hat{\mathcal{S}}_+ &=&
e^{-{\rm i} \hat p} \big[ s(s+1)- \hat x(\hat x+1) \big]^{1/2} , \\
\hat S_- \to \hat{\mathcal{S}}_- &=&
 \big[ s(s+1)- \hat x(\hat x+1) \big]^{1/2} e^{{\rm i} \hat p} , \\
 \hat S_0 \to \hat{\mathcal{S}}_0 &=& x .\eeqa
Thus, it would appear likely that the SHA Hamiltonian (\ref{eq:su2_sha_limit})
could be obtained as a contraction of this realisation.

\subsection{SU(1,1){CG } coefficients}

We now compare the complementary RPA and SHA derivations of asymptotic SU(1,1) CG coefficients
by diagonalization of a Hamiltonian
\beq \hat H = \alpha\hat K_0  + \chi \hat K_+ \hat K_- .
\label{Hsu11approx}\eeq
The spectrum of eigenstates of this Hamiltonian are the SU(1,1) states
$\{ |Kn\rangle\}$ with eigenvalues given by
\beq E_{Kn} = \alpha (K+n) +  \chi (2K +n-1)n ,\eeq
or, with a relabelling of states by $|Kn\rangle \to |Nn\rangle$, where
$N=K+n-k_1-k_2$, by
\beq  \mathcal{E}_{Nn} =  \alpha (k_1+k_2+N) +  \chi (2k_1+2k_2+2N-n -1)n .
\label{eq:E_Nn}\eeq
For $\alpha$ and $\chi$ positive, the lowest value of $ \mathcal{E}_{Nn} =E_{Kn}$ is for $n=N=0$.
Then, for asymptotically large values of $k_1$ and $k_2$,  the low-lying eigenvalues are given to leading order in $N$ and $n$  by
\beq  \mathcal{E}_{Nn} \sim  \alpha (k_1+k_2) + \alpha N + \chi (2k_1+2k_2-1)n ,\eeq
which are the eigenvalues of the Hamiltonian
\beq \hat H \sim  \alpha (k_1+k_2) +  \alpha c^\dag c +
\chi (2k_1+2k_2-1) b^\dag b \label{eq:RPAHamsu11a}\eeq
 for a two harmonic oscillator system with
\beq c^\dag c |Nn\rangle = N |Nn\rangle , \quad
b^\dag b |Nn\rangle = n |Nn\rangle .\eeq

The corresponding eigenvectors, in the space of coupled tensor product states
$\{ |k_1n_1\rangle \otimes  |k_2n_2\rangle \}$, and hence a subset of asymptotic SU(1,1) CG coefficients, can be obtained in the RPA by a contraction of the two su(1,1) irreps $\{ k_1\}$ and $\{ k_2\}$.
The relevant contraction is obtained by observing that for small values of $n$, the  su(1,1) relationships
\beq  [\hat K_-, \hat K_+] |kn\rangle = 2\hat K_0 |kn\rangle  = 2(k+n) |kn\rangle, \quad
[\hat S_0, \hat S_{\pm}] = \pm \hat S_\pm , \eeq
are satisfied in the $k\to \infty$ asymptotic limit, for finite values of $n$, by
\beq \frac{1}{\sqrt{2k}} \,\hat K_+ \sim  a^\dag , \quad
\frac{1}{\sqrt{2k}} \, \hat K_- \sim  \, a, \quad
 \hat K_0 \sim k + a^\dag a . \label{eq:su11contraction} \eeq
Thus, by expressing the Hamiltonian  of Eqn.\ (\ref{Hsu11approx}) in terms of the su(1,1) operators of the $\{ k_1\}$ and $\{ k_2\}$ sub-representations, it becomes the Hamiltonian for two coupled harmonic oscillators which, in the RPA, are decoupled by diagonalization to give the uncoupled oscillator expression of Eqn.\ (\ref{eq:RPAHamsu11a}).
 The contraction (\ref{eq:su11contraction}) which leads to this RPA result is valid in the domain of coupled states, $\{ |Kn\rangle\}$, in which $K$ is close to $k_1+k_2$ and
 $n$ is small.

Other su(1,1) CG coefficients are obtained by considering the eigenvectors of the Hamiltonian (\ref{Hsu11approx}) in the subspace of states of a fixed value of $N$.
For asymptotically large $N$ and small $n$, Eqn.\ (\ref{eq:E_Nn}) then reduces to the eigenvalues
\beq  \mathcal{E}_{Nn} \sim  \alpha (k_1+k_2+N) + 2 \chi (k_1+k_2+N)n
\eeq
of the Hamiltonian
\beq \hat H \sim  \alpha (k_1+k_2+N) +
2\chi (k_1+k_2+N) b^\dag b \label{eq:SHAHamsu11a}\eeq
for a simple harmonic oscillator with $b^\dag b |Nn\rangle = n |Nn\rangle$.
The SHA gives an explicit expression for this Hamiltonian as a shifted harmonic oscillator Hamiltonian.  We have not succeeded in deriving this expression in terms of a contraction of the su(1,1) algebra.
However, the SHA is based on an approximation to a realisation of the su(1,1) Lie algebra given by the expressions, in terms of the harmonic oscillator operators
$\hat x=x$, $\hat p= -{\rm i}\hbar d/dx$, 
\beqa \hat K_+ \to \hat{\mathcal{K}}_+ &=&
e^{-{\rm i} \hat p} \big[(2k+\hat x)(\hat x+1) \big]^{1/2} , \\
\hat K_- \to \hat{\mathcal{K}}_- &=&
 \big[ (2k+\hat x)(\hat x+1) \big]^{1/2} e^{{\rm i} \hat p} , \\
 \hat K_0 \to \hat{\mathcal{K}}_0 &=& k+\hat x .\eeqa
Thus, it would appear once again  likely that the Hamiltonian (\ref{eq:SHAHamsu11a})
can be obtained as a contraction of this realisation.

\section{Discussion and conclusion}

In this paper we have demonstrated that turning a three-term recursion relation into a
second order differential equation makes it easy to understand the oscillatory nature of CG coefficients.
 From a more numerical perspective, it has been noted that the final forms
given in the paper, even the simplified expressions, are remarkably accurate even for values of the parameters that are
far from the asymptotic limits in which they become precise.

It is readily ascertained that
our asymptotic
 SU(2) CG coefficients retains the symmetry
\beq ( s_{1}m_1\, s_{2}m_2 |s_{1}\!+\!s_{2}\!-n,M) =
(-1)^n ( s_{2}m_2\, s_{1}m_2 |s_{1}\!+\!s_{2}\!-n,M) \eeq
 under the exchange $s_1m_1 \leftrightarrow s_2m_2$.
 Indeed, this is seen from the simplified expression of Eq.\ (\ref{mainsu2result2}) in which the argument of the Hermite polynomial simply changes its sign under this exchange.  A parallel result is obtained for an SU(1,1) CG coefficient for which, in the sign convention used,
 \beq C( k_{1},n_1, k_{2},n_2 ; K,n) =
(-1)^n  C( k_{2},n_2, k_{1},n_1 ; K,n) .\eeq
Due to the   asymmetrical  way in which the asymptotic limits are approached, it is not expected that the more general symmetries of these CG coefficients will be preserved. However, such symmetries as are known, e.g., for the exchange
$s_1m_1 \leftrightarrow s_3,-m_3$, can be used to re-arrange the arguments of a given CG coefficient such that the value of $n$ is minimised. The effective range over which the asymptotic approximations are expected to produce acceptable results is thereby increased.

More generally, the successes of the SHA in its applications to date
\cite{Chen,SHAIBM,SYHo} suggest that it is
a potentially powerful technique that could be applied more generally.
Note, for example, that the finite and discrete series of irreps of all semi-simple Lie algebras are characterized by the irreps of their many su(2)
and/or su(1,1) subalgebras generated by raising and lowering operators.
In particular, every pair of raising and lowering operators, $X^\pm_\nu$,
of a semi-simple Lie algebra
can be normalized to have
 SU(2) commutation relations
$[X^+_\nu, X^-_\nu] = h_\nu$, $[h_\nu , X^\pm] = \pm 2 X^\pm_\nu$.
Note also that many dynamical systems have low-energy collective states that are well approximated as harmonic vibrational states.  As a result the RPA (random phase approximation) has become a powerful tool in many-body theory.
It is also of interest to explore the possibility of mapping the algebraic structure of a problem to a rotor algebra, for example, rather than that of a harmonic vibrator.

The search for an extension of the SHA to apply to many-body systems with more general algebraic structures is worthwhile because it is known that, for a variety of many-body systems, the RPA works well for describing vibrational normal mode excitations with relatively weak interactions but breaks down when the interactions become too strongly attractive.  This is the situation in which the system is understood to undergo a phase change; the frequency of one of the vibrational modes goes to zero  and a deformed equilibrium state, with rotational plus vibrational degrees of freedom, emerges.  Thus, we are optimistic that a generalised SHA will prove to be appropriate for such situations.
Section 4 has shown the SHA to be  complementary to the RPA its ability to give (asymptotic) solutions to an eigenvalue equation in regions where the RPA breaks down.  Moreover, 
the SHA has been shown, in previous applications, to provide practical solutions of the multi-level BCS Hamiltonian that are remarkably accurate for relatively strong pairing interactions \cite{SYHo}.

\black

\appendix
\section{A special case}

The special case
\begin{equation}
(s_1,-m;s_2,m|s_1+s_2,0)\sim \left(\frac{1}{\pi}\left(\frac{s_1+s_2}{s_1s_2}\right)
\right)^{\frac{1}{4}}\,e^{-m^2 (s_1+s_2)/2s_1s_2} \label{specialcaseapprox}
\end{equation}
follows by careful application of Stirling's formula
\beq
  s!\sim \sqrt{\frac{2\pi}{s}}\, s^{s+1} e^{-s} \, , \label{Stirling}
\end{equation}
to the exact expression
\beqa
&&(s_1,-m,s_2,m| s_1+s_2,0) \nonumber \\
&&\qquad\qquad = \left[
\frac{(2s_1)! (2s_2)! (s_1+s_2)! (s_1+s_2)!}
{(2s_1+2s_2)! (s_1+m)!(s_1-m)! (s_2+m)! (s_2-m)!} \right]^{\hf} .\label{specialcaseexact}
\eeqa

First, one can use Eqn.(\ref{Stirling}) to show that
\beq
\frac{s! s!}{(s+m)!(s-m)!} \sim \sqrt{\frac{s^2-m^2}{s^2}}\,
\frac{s^{2s+2}}{(s+m)^{s+m+1} (s-m)^{s-m+1}} .
\eeq
With $x=m/s$, this expression reduces to
\beq
\frac{s! s!}{(s+m)!(s-m)!} \sim \frac{1}{\sqrt{1-x^2}}\,
\left[ \frac{(1-x)^x}{(1+x)^x} \frac{1}{(1-x^2)} \right]^s
\eeq
and, for a finite value of $m$, reduces further in the limit as $s\to\infty$ to
\beq
\frac{s! s!}{(s+m)!(s-m)!} \sim \frac{1} {(1+m^2/s^2)^s} = e^{-m^2/s} .
\eeq
Eqn.(\ref{specialcaseexact}) can then be manipulated to directly yield Eqn.(\ref{specialcaseapprox}).

As a special case we observe that
\beqa
   (s,-m;s,m|2s,0) &=&  \frac{(2s)! (2s)! }{\sqrt{(4s)!}\, (s+m)!(s-m)!} \, , \\
   &=& \left(\frac{2}{\pi s}\right)^{\frac{1}{4}}\,e^{-m^2/s}\, .\label{specialspecialcase}
\eeqa

Both Eqn.(\ref{specialcaseapprox}) and Eqn.(\ref{specialspecialcase}) agree with Eqn.(\ref{mainsu2result}) for $n=0$ and $M=0$ in the limit where $s_1,s_2\to\infty$.


\begin{thebibliography}{99}
\bibitem{su11code} Exact su(1,1) CG coefficients were calculated with a Maple code supplied by J.~Van der Jeugt.
\end{thebibliography}
\end{document}